\documentclass[numreferences]{crckbked}

\makeindex
\usepackage{times}
\usepackage{bm}
\usepackage{klucite}
\usepackage{graphicx}

\begin{document}

\begin{opening}
\title{Influence of thermal fluctuations on an
underdamped Joseph\protect\-son tunnel junction}

\author{I.S. \surname{Beloborodov}$^{1,2}$}
\author{F.W.J. \surname{Hekking}$^3$}
\author{F. Pistolesi$^3$}

\institute{$^1$Bell Laboratories, Lucent Technologies\\
Murray Hill, New Jersey 07974\\
$^2$Department of Physics, University of Colorado\\
CB 390, Boulder, Colorado 80390\\
$^3$Laboratoire de Physique et Mod\'elisation des Milieux Condens\'es\\
Magist\`ere-CNRS , BP 166, 38042 Grenoble-cedex 9, France}

\runningtitle{Thermal fluctuations in a Josephson junction} \runningauthor{I.S.
Beloborodov, F.W.J. Hekking, and F. Pistolesi}

\begin{abstract}
Inspired by a recent experiment, we study the influence of thermal
fluctuations on the $I$-$V$ characteristics of a Josephson junction,
coupled to a strongly resistive environment. We obtain analytical
results in the limit where the Josephson energy is larger than the
charging energy and quasiparticles are absent.
\end{abstract}

\end{opening}

\section{\label{Introduction}Introduction}

As it is well-known, the dynamics of a small Josephson tunnel junction is
characterized by two energies: the Josephson coupling energy $E_J$ and the charging
energy $E_C$~\cite{Tinkham96}. The behavior of the junction is determined by the
competition between these two energy scales. If the Josephson coupling energy
dominates, a superconducting state with a well-defined phase difference $\phi$ across
the junction is possible. The junction will carry a Cooper pair current $I=I_J \sin
\phi$ in the absence of an external voltage $V$. Here $I_J = 2e E_J/\hbar$ is the
Josephson critical current ($e$ is the electron charge). If on the other hand the
charging energy dominates, an insulating state with a well-defined charge $Q$ on the
junction is possible. This gives rise to a gap in the $I$-$V$ characteristics of the
junction, associated to Coulomb blockade~\cite{Grabert90}: the Cooper pair current
$I=0$ up to a voltage $V = 2 E_C/e$. Here we defined $E_C = e^2/2C$, where $C$ is the
capacitance of the junction.

\begin{figure}[h]
\centerline{\includegraphics[width=0.45\textwidth]{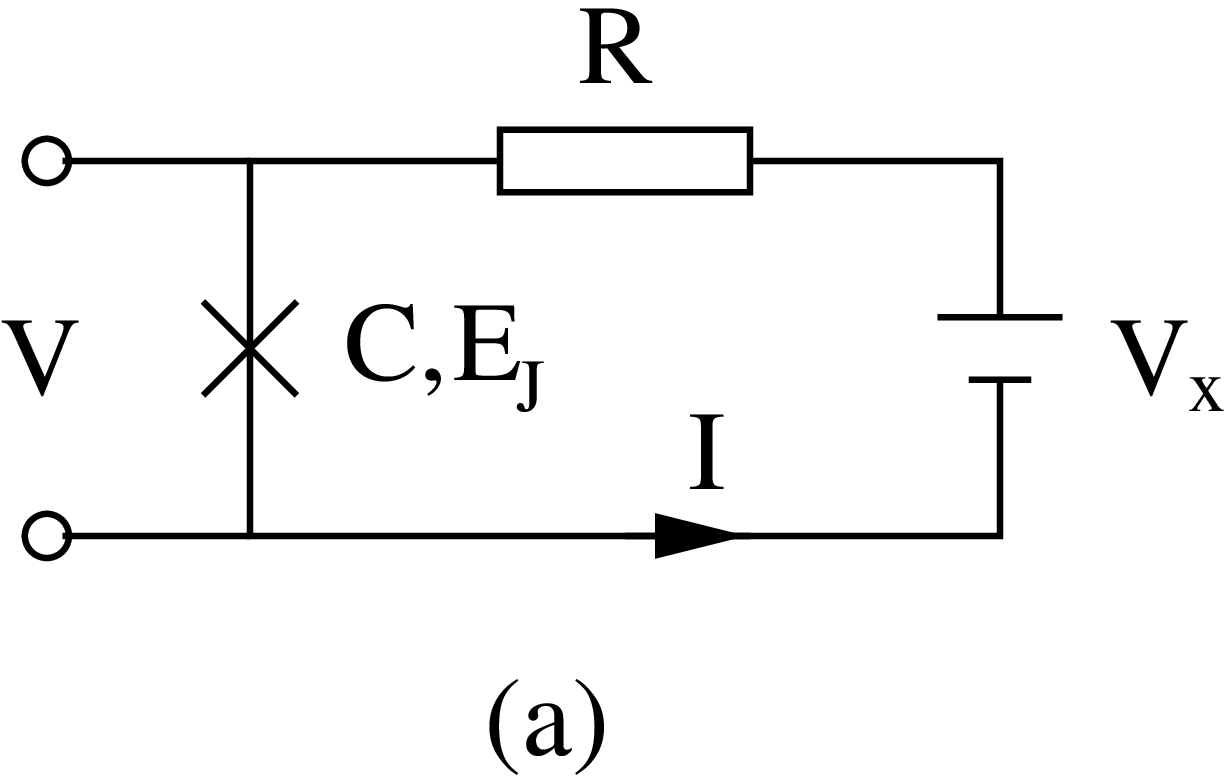}\hspace{1cm}\includegraphics[width=0.45\textwidth]{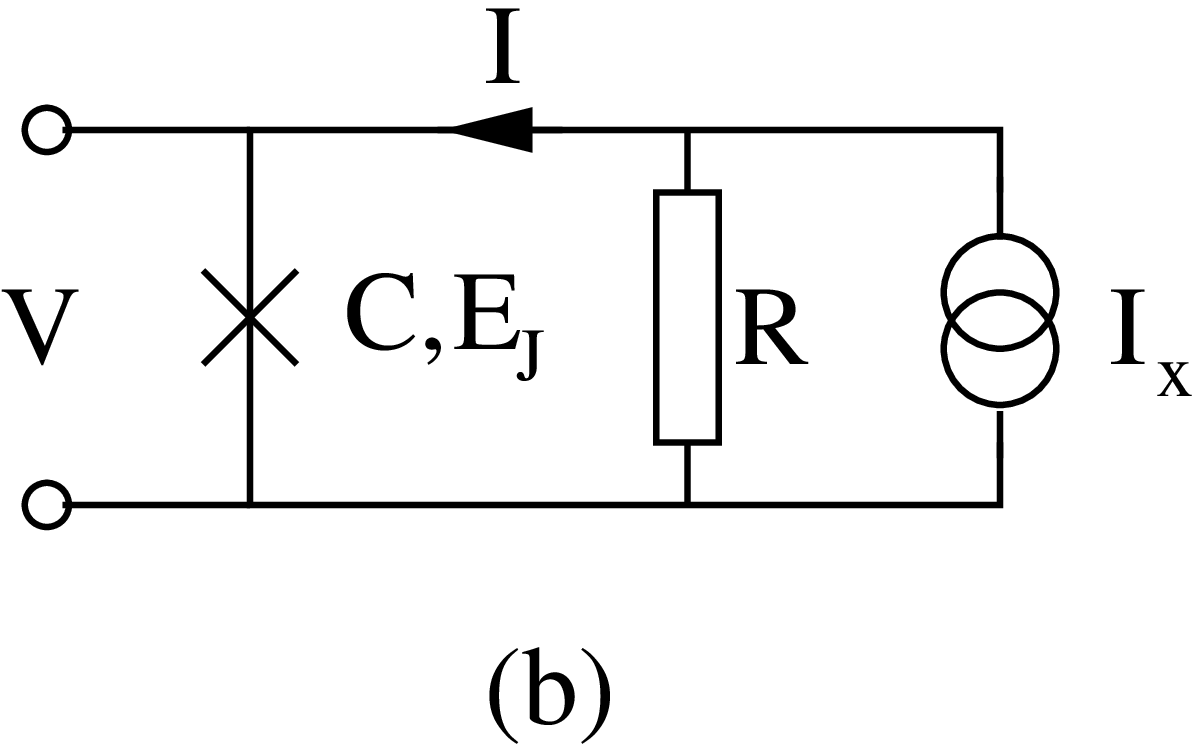}}
\vspace{2.5 mm} \caption{Voltage biased (a) and current-biased (b) Josephson
junction.} \label{setup}
\end{figure}

The above statements are true under appropriate conditions on the impedance
$Z(\omega)$ of the circuit connected to the junction, as we will discuss in some
detail below. To be specific, we will consider two configurations. The voltage-biased
set-up, in Fig.~\ref{setup}a, consists of an ideal voltage source (voltage $V_x$),
connected to a series arrangement of a Josephson junction and a resistance $R$. In a
current-biased set-up, see Fig.~\ref{setup}b, an ideal current source (bias current
$I_x$) is connected to a parallel arrangement of a Josephson junction and a shunt
resistance $R$. The two set-ups are equivalent if we choose $I_x =
V_x/R$~\cite{Ingold90}. The resistance $R$ plays a crucial role and determines the
detailed form of the $I$-$V$ characteristics. In particular, the way in which the
transition from superconducting to insulating behavior of the junction manifests
itself in the current-voltage characteristic depends not only on the parameters $E_J$
and $E_C$, but also on $R$.

The voltage biased set-up, Fig.~\ref{setup}a, has been analyzed long
ago by Ivanchenko and Zil'berman~\cite{Ivanchenko68} in the limit of
strong charging effects, $E_C \to \infty$. More recently, a
perturbative approach has been developed to obtain the current-voltage
characteristic~\cite{Ingold90,Ingold94}. This approach is valid for
finite, but small, values of $E_J/E_C$. Specifically, in the zero
temperature limit $T=0$, a change of the position of the supercurrent
peak is found upon increasing the ratio $R/R_Q$, where $R_Q =
h/4e^2$. For $R/R_Q < 1$, the supercurrent peak is centered at zero
bias voltage; its width increases gradually as $R/R_Q$ is increased.
If $R/R_Q \gg 1$, the supercurrent peak is centered around $eV = 2
E_C$. This corresponds to the transition from superconducting behavior
found for small values of the external resistance to a complete
Coulomb blockade of Cooper pair tunneling in the presence of a
resistive environment. The range of applicability of the perturbative
analysis depends on $E_J$, $E_C$, and $R$. In the low resistance limit
$R/R_Q < 1$, the approach breaks down at low energy (bias voltage or
temperature) for any value of $E_J$. In the opposite limit $R> R_Q$,
the condition $E_J/E_C \ll (R_Q/R)^{1/2}$ needs to be satisfied. A
non-perturbative analysis was developed in~\cite{Ingold99}, however
its validity is limited to low energies, $eV ,k_BT \ll E_C R_Q/R$.

The approach discussed so far is clearly appropriate if one is interested in a small
junction with $E_J \leq E_C$, coupled to an environment which is not too resistive. An
example is the recent experiment by Steinbach {\em et al.}~\cite{Steinbach01}, in good
agreement with the theoretical predictions of Refs.~\cite{Ivanchenko68,Ingold99}.
However, the perturbative approach is not applicable in the case of a Josephson
junction with $E_J \ge E_C$, coupled to a strongly resistive environment, $R \gg R_Q$,
usually referred to as the underdamped limit~\cite{Tinkham96}. Nevertheless this is a
relevant situation from an experimental point of view as well. Recently, Watanabe and
Haviland measured the $I$-$V$ characteristics of a small Josephson junction coupled to
a tunable environment~\cite{Watanabe01}. In these experiments, $E_J \sim E_C$ and the
environment could be tuned to large values of the resistance such that $R/R_Q \sim
10^2$ -- $10^4$. No evidence of a supercurrent peak was observed; rather a so-called
Bloch nose~\cite{ALZ85} was found for small values of the current and the bias
voltage. This corresponds to an on-set of Coulomb blockade at low voltage and small
current, followed by subsequent back-bending of the $I$-$V$ curve to smaller values of
the voltage at higher currents which indicates a cross-over to superconducting
behavior. The shape of the Bloch nose depended on the resistance $R$ as well as on
temperature. The experimental findings could not be reconciled quantitatively with
existing theories of Bloch nose~\cite{ALZ85,Schoen90}. The main problem is that those
theories are essentially based on the presence of dissipation due to quasiparticles.
In the experiment~\cite{Watanabe01}, the tunable environment is a long array of
superconducting junctions in the insulating regime~\cite{Fazio01}, and quasiparticles
are practically absent.

In this paper we analyze the finite temperature behavior of a Josephson junction with
$E_J \gg E_C$, coupled to a very resistive environment, in the absence of
quasiparticles. We obtain, to our knowledge for the first time, explicit formul{\ae}
for the current-voltage characteristics at finite temperature for a current-biased
configuration with large $R$, see Fig.~\ref{setup}b. We follow the approach pioneered
in Ref.~\cite{ALZ85}, which is based on the periodicity of the Josephson coupling
energy as a function of the phase difference $\phi$ across the junction. As a result
of this periodicity, the eigenstates of the junction are of the Bloch type, and the
energy spectrum consists of bands of width of the order $E_C$, separated by gaps of
the order $E_J$. In the limit $E_J \gg E_C$, the gaps are large and the junction stays
in the lowest Bloch band at low temperature $k_BT \ll E_J$, as long as the bias
current $I_x \ll I_J$. If $R=\infty$, Bloch oscillations~\cite{ALZ85} of the voltage
$V$ occur, such that its time-averaged value vanishes: the junction is in the
superconducting state. If $R$ is finite, relaxation within the lowest band modifies
the dynamics of the junction. This results in the appearance of the Bloch nose: if
$I_x$ is so small that relaxation prevents Bloch oscillations to occur, a finite
voltage state develops with no current passing through the junction (Coulomb
blockade), all current passes through the resistance. If $I_x$ is increased above a
threshold value, Bloch oscillations develop and the a superconducting zero-voltage
state is reached. Our approach is valid for a large temperature range and our results
can be compared quantitatively with the experimental data of Ref.~\cite{Watanabe01}.

\section{The model}
\label{model}

The circuit of Fig.~\ref{setup}b can be modelled by the Hamiltonian $ \hat{H} =
\hat{H}_0 + \hat{H}_\mathrm{coupl} + \hat{H}_\mathrm{env}$. Here,
\begin{equation}
\hat{H}_0 = \frac{\hat{Q}^2}{2C} - E_J \cos \hat{\phi} \label{H0}
\end{equation}
describes the Josephson junction in the absence of the external circuit. The charge
$\hat{Q}$ and the phase $\hat{\phi}$ are canonically conjugate operators,
$[\hat{Q},\hat{\phi}] = 2ie$. Since the Hamiltonian~(\ref{H0}) is periodic in $\phi$,
the energy spectrum consists of bands $\epsilon_n(q)$ where $n$ is the band index and
$q$ the quasimomentum associated to $\phi$, {\em i.e.}, the quasicharge. The spectrum
$\epsilon_n(q)$ is a periodic function of $q$ with period $2e$. We will limit
ourselves to the first Brillouin zone, $-e \le q \le e$. The corresponding eigenstates
$\psi _{nq}(\phi)$ are of the Bloch type. In the limit of interest here, $E_J \gg
E_C$, the bands are narrow compared to the gaps. Throughout this paper we will assume
that all relevant energies are smaller than $E_J$ and restrict ourselves to the lowest
energy band, $n=0$, with a dispersion $\epsilon_0(q) = \hbar \sqrt {8 E_C E_J} - U_0
\cos (\pi q/e)$, where the bandwidth is given by
\begin{equation}
U_0 = 16 \sqrt{\frac{2}{\pi}}E_C[\frac{E_J}{2E_C}]^{3/4}e^{-\sqrt{8E_J/E_C}} .
\label{U0}
\end{equation}

The term
\begin{equation}
\hat{H}_\mathrm{coupl} = -  \frac{\hbar}{2e}\hat{I} \hat{\phi} \label{Hcoupl}
\end{equation}
couples $\hat{\phi}$ to the total current $\hat{I}= I_x + \delta\hat{I}$, which
contains the constant bias current $I_x$ and a fluctuating part $\delta\hat{I}$.  The
fluctuations are induced by the environment (the resistance $R$ in Fig.~\ref{setup}b)
and the dynamics of $\delta \hat{I}$ is governed by $\hat{H}_\mathrm{env}$.

In the absence of any current, $\hat{H}_\mathrm{coupl} =0$ and
quasicharge $q$ is a well defined variable. In the limit of small bias
current $I_x$ and current fluctuations $\delta I$, the effect of the
coupling term~(\ref{Hcoupl}) on $q$ can be analyzed perturbatively. As
it was shown in Refs.~\cite{ALZ85}, a quantum Langevin
equation for the quasicharge operator $\hat{q}$ can be obtained, which
reads
\begin{equation}
\frac{d\hat{q}}{dt} = I_x - \frac{1}{R}\frac{d\epsilon _0}{d\hat{q}} +
\delta \hat{I}.
\label{quantlang}
\end{equation}
Here, $I_x$ plays the role of a driving force and $\delta \hat{I}$ of
a random force.  The statistical properties of the latter are
determined by the Hamiltonian $\hat{H}_\mathrm{env}$. Specifically,
for a resistive environment at equilibrium we have~\cite{Landau}
\begin{equation}
\langle \delta \hat{I} (t) \rangle_\mathrm{env} = 0 \; \; , \langle
[\delta \hat{I}(t),\delta \hat{I}(t')]_+ \rangle
_{\mathrm{env},\omega} = 2 (\hbar \omega/R) \coth [\hbar
\omega/(2k_BT)] \label{stat},
\end{equation}
where $[\ldots,\ldots]_+$ denotes the anticommutator and $\langle
\ldots \rangle_\mathrm{env}$ an average with respect to
$\hat{H}_\mathrm{env}$. Finally, the derivative $d\epsilon
_0/d\hat{q}$ is the voltage operator $\hat{V}$ across the junction,
the corresponding term $V/R$ acts as a damping term.

We will be particularly interested in the limit of small current
fluctuations. In this limit, quasicharge remains a well defined
quantity and we can replace operators by classical variables in
Eq.~(\ref{quantlang}). In the next Section, we will analyze the
resulting classical Langevin equation in some detail, using well-known
methods for stochastic equations~\cite{vanKampen}. We will see under
which condition current fluctuations can be considered small and we
will obtain the detailed, analytical form of $I$-$V$ characteristics,
as a function of the resistance $R$ and of temperature $T$.

\section{Current-voltage characteristics}
\label{I-V}

\subsection{Linearized Langevin equation}
\label{langevin} In the absence of fluctuations, $\delta\hat{I} =0$, the Langevin
equation~(\ref{quantlang}) has a stationary solution when $I_x = (1/R)
d\epsilon _0/dq \equiv V/R$, with
$q=q_0=(e/\pi)\arcsin(I_x/I_b)$. Here, $I_b=V_b/R$, with $V_b = \pi
U_0/e$ the maximum voltage across the junction corresponding to the
maximum slope of $\epsilon _0 (q)$. The corresponding $I$-$V$
characteristic is linear, $I_x = V/R$. In other words, all the current
passes through the resistance $R$, no current flows through the
junction (Coulomb blockade) and the phase $\phi$ is completely
undefined.  This is consistent with the fact that we treat $q$ as a
well-defined variable. Note that the stationary solution exists only
as long as $I_x\leq I_b$; for larger current, $I_x > I_b$, the
quasicharge becomes time-dependent. We will discuss that case in
Sec.~\ref{Fokker-Planck} below.

Let us now take into account small fluctuations $\delta\hat{I}$ such
that the induced charge fluctuations $\delta q$ around the stationary
solution $q=q_0$ are small. We consider the linear part of the average
current-voltage characteristics and calculate
\begin{equation}
\label{iv} \langle V\rangle= \langle d\epsilon_0/dq \rangle = V_b\langle \sin(\pi q/e)
\rangle.
\end{equation}
Here the symbol $\langle \ldots\rangle$ means averaging over the
fluctuations of charge. In order to find the distribution of these
fluctuations, we write $q=q_0 + \delta q$, such that the linearized,
classical Langevin equation takes the form
\begin{equation}
\label{fluctuations} \delta \dot{q} = -\frac{\delta q}{\tau}+ \delta I,
\label{linlang}
\end{equation}
where we introduced the relaxation time $\tau$ such that $\tau^{-1} =
(\pi/e) \sqrt{|I_b^2-I_x^2|}$. The time $\tau$ defines the time scale
characterizing the junction dynamics. Note that this is a relatively
long time compared to the $RC$ time of the circuit: for $I_x =0$, we
find $\tau \sim e/I_b \sim R e^2 / U_0 \gg RC$ as $U_0$ is
exponentially small, see Eq.~(\ref{U0}).

In order to solve the stochastic equation~(\ref{linlang}), we should
specify the correlation function $\langle \delta I(t) \delta
I(t')\rangle$, see Eq.~(\ref{stat}).  The frequencies $\omega$ of
interest here are small, $\omega \sim 1/\tau$, determined by the long
time scale $\tau$ characterizing the dynamics of quasicharge. For
finite temperatures $k_BT \gg \hbar /\tau$, the correlation function
$\langle \delta I(t) \delta I(t')\rangle _\omega \simeq 2 k_BT/R$,
independent of frequency and thus $\langle \delta I(t) \delta
I(t')\rangle = (2k_B T/R) \delta (t-t')$. This means that the relevant
current fluctuations are classical and completely uncorrelated on the
long time scale $\tau$ characterizing the quasicharge dynamics.

The stochastic equation~(\ref{linlang}) can now be solved, using
standard methods developed for the analysis of Brownian
motion~\cite{vanKampen}. As a result, one obtains the probability
distribution $W(\delta q,t;\delta q_0)$ to find the fluctuation
$\delta q$ at time $t$ given that $\delta q = \delta q_0$ at time
$t=0$.  In the long time limit, $t \rightarrow \infty$, this
probability distribution does not depend on time and is independent of
$\delta q_0$. It is given by the Gaussian distribution
\begin{equation}
\label{Gauss} W(\delta q, \tau)= \sqrt{\frac{1}{\pi \gamma \tau}}
\exp\left(-\frac{(\delta q)^2}{\gamma \tau}\right),
\end{equation}
where we introduced the parameter $\gamma=2k_B T/R$.

Using Eq.~(\ref{Gauss}) we immediately obtain the average square of
the charge fluctuations, $\langle (\delta q)^2 \rangle = k_B T \tau
/R$. Quasicharge can be considered well-defined as long as these
fluctuations are small, $\langle (\delta q)^2 \rangle \ll e^2$. This
condition gives an additional restriction on temperature, $k_B T \tau
/\hbar \ll R/R_Q$. Together with the condition $k_BT \gg \hbar /\tau$
found above we obtain the temperature window
\begin{equation}
1 \ll k_B T\tau /\hbar \ll R/R_Q .\label{Tint}
\end{equation}
This means in particular that $R/R_Q \gg 1$ for the analysis presented here to be
correct. If this condition is verified, quasicharge is a well-defined variable, and
quantum fluctuations can be neglected.  For small $I_x$ the energy scale $\hbar /
\tau$ is of the order of $eV_b R_Q/R$, thus the condition~(\ref{Tint}) reads $eV_b
R_Q/R \ll k_B T \ll e V_b$. This has a transparent physical interpretation: the
temperature has to be much smaller than $eV_b$ in order not to smear the Bloch nose,
but it must be larger than $e V_b (R_Q/R) \ll eV_b$ in order to justify the neglect of
quantum fluctuations. The experiment of Ref.~\cite{Watanabe01} was performed at very
large values of $R/R_Q$, such that the condition~(\ref{Tint}) was verified.

With the help of Eqs.~(\ref{iv}) and (\ref{Gauss}) we readily obtain
the average voltage for the linear part of the average $I$-$V$
characteristic,
\begin{equation}
\label{result1} \langle V \rangle = I_x R \left[ 1 - \frac{\pi}{2} \frac{k_B T}{e R}
\frac{1}{\sqrt{I_b^2-I_x^2}} \right].
\end{equation}
Equation~(\ref{result1}) represents the main result of this
section. The first term on the right hand side of Eq.~(\ref{result1})
coincides with the linear $I$-$V$ curve which was obtained above
without taking into account current fluctuations. The second term is
entirely due to the thermal fluctuations, which suppress the
resistance. In other words, the Coulomb blockade is smeared due to
thermal activation.

In this subsection we have considered the limit of small enough bias
current, $I_x < I_b$, such that the Langevin equation could be
linearized around a well-defined, stationary solution. For larger
values of $I_x$, no stationary solution exists: quasicharge is an
oscillating function of time and the full non-linearity of the
Langevin equation should be taken into account. In
Section~\ref{Fokker-Planck} we will use a Fokker-Planck approach to
obtain the complete current-voltage characteristics.

\subsection{Fokker-Planck approach}
\label{Fokker-Planck}

If the applied bias current $I_x$ exceeds $I_b$, quasicharge is a
dynamical variable even in the absence of fluctuations. The
corresponding non-linear equation of motion can be integrated, though,
and $q(t)$ can be found. As a result, the voltage $V$ becomes an
oscillating function of time, with a period given by $\tau$. Its
time-averaged value $\bar{V}$ over one period of the oscillation is
given by~\cite{Schoen90}
\begin{equation}
\bar{V} = I_b R\left[\frac{I_x}{I_b}-\sqrt{(I_x/I_b)^2-1} \right]. \label{V0}
\end{equation}
In other words, on measurement time scales much longer than the period $\tau$, the
voltage oscillations average out and a stationary situation is reached.

In order to investigate the influence of small current fluctuations
$\delta I$ on the $I$-$V$ characteristics, the full non-linear
stochastic problem~(\ref{quantlang}) needs to be solved. This can be
conveniently done using a Fokker-Planck approach~\cite{vanKampen}. In
this approach, the probability distribution $W(q,t)$ to find
quasicharge $q$ at time $t$ is found , as a solution of the
Fokker-Planck equation~\cite{vanKampen}
\begin{equation}
\label{Fokker} \frac{\partial W}{\partial t} = - \frac{\partial}{\partial
q}\left(\left[I_x-I_b \sin\left(\frac{\pi q}{e}\right)\right]W \right) +
\frac{\gamma}{2}\frac{\partial^2 W}{\partial q^2}.
\end{equation}
In writing this equation, we limited ourselves again to the case of classical
fluctuations only, $k_BT \gg \hbar/\tau$. In addition, we note that this equation can
be used only as long as the fluctuations are $\delta$-correlated on the time scale
defined by $\tau$. Hence, we need to restrict ourselves again to the temperature
interval~(\ref{Tint}).

We are particularly interested in the stationary solution
of~(\ref{Fokker}), $W(q)$, reached in the long time limit $t \gg
\tau$. The average voltage is then given by
\begin{equation}
\label{V} \langle V\rangle = I_bR \int\limits_{-e}^{+e}dq  \sin(\pi q/e) W(q),
\end{equation}
where $W(q)$ is normalized, $\int_{-e}^{+e} W(q)dq = 1$. In the long
time limit $t \gg \tau$, this corresponds to the measured,
time-averaged voltage, $\langle V \rangle = \bar{V}$.

\begin{figure}[h]
\centerline{\includegraphics[width=0.45\textwidth]{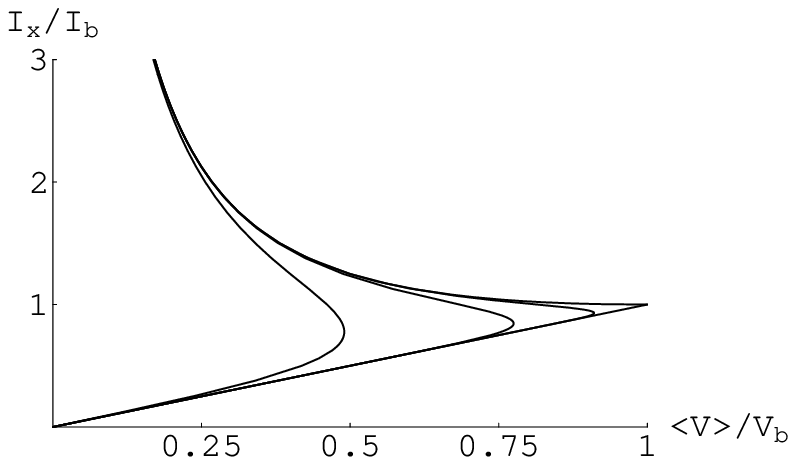}\includegraphics[width=0.45\textwidth]{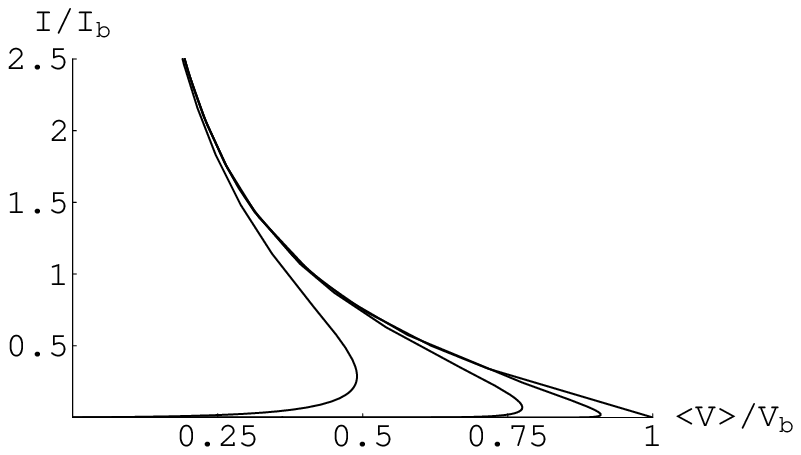}}
\vspace{2.5 mm} \caption{Calculated $I$-$V$ characteristics for an underdamped
junction. Left panel: bias current $I_x/I_b$ as a function of voltage $\langle V
\rangle/V_b$ across the junction, for various values of the parameter $k_BT/eV_b$.
Curves from right to left correspond to $k_BT/eV_b =$ 0, 0.004, 0.02, and 0.1. Right
panel: current $I/I_b$ through the junction as a function of $\langle V \rangle/V_b$.
Curves from right to left correspond to the same values of $k_BT/eV_b$.}
\label{IVcurve}
\end{figure}

Let us first analyze the solution of the problem in the absence of
charge fluctuations [parameter $\gamma $ equals to zero in
Eq.~(\ref{Fokker})]. Solving Eq.~(\ref{Fokker}) and using the
normalization condition for the distribution function we obtain
\begin{eqnarray}
\label{p} W(q)= \left \{
\begin{array}{lr}
\delta(q-q_0), & I_x\leq I_b \\ \frac{\sqrt{(I_x/I_b)^2-1}}{2e [I_x/I_b-\sin(\pi
q/e)]}, & I_x > I_b,
\end{array}
\right.
\end{eqnarray}
where $q_0$ is the stationary solution found in Section~\ref{langevin}. Substituting
Eq.~(\ref{p}) into Eq.~(\ref{V}) for the current-voltage characteristics we
immediately obtain the linear result $\langle V \rangle = I_x R$ for $I< I_b$; for the
opposite case $I_x > I_b$ we reproduce the result~(\ref{V0}). The resulting $I$-$V$
characteristic, a so-called Bloch nose, is shown in Fig.~\ref{IVcurve} as the curve
corresponding to the parameter $k_BT/eV_b = 0$.

Next we consider the current-voltage characteristics in the presence of fluctuations.
To find the distribution function, $W(q)$, we use a Fourier representation
\begin{equation}
\label{fourie} W(q)=\sum\limits_{n=-\infty}^{+\infty}W_n e^{i \pi qn/e}.
\end{equation}
Substituting the right hand side of Eq.~(\ref{fourie}) into
Eq.~(\ref{Fokker}) and using the periodicity condition,
$W(q+2e)=W(q)$, we obtain the following recurrence relation for the
distribution function
\begin{equation}
\label{recurent} 2\left(n + i\eta_x \right)W_n = \eta_b\left(W_{n-1} - W_{n+1}\right).
\end{equation}
Here we introduced the short hand notation $\eta_j = 2eI_j/(\pi \gamma)$ for $j=x,b$.
To find the distribution function $W_n$ in Eq.~(\ref{recurent}) we use the analogy
with the recurrence relation for the modified Bessel function~\cite{Gradstein} $2\nu
I_{\nu}(\xi) = \xi[I_{\nu-1}(\xi) - I_{\nu+1}(\xi)]$. Here $\nu$ is the index and
$\xi$ is the argument of the modified Bessel function. A comparison
with~(\ref{recurent}) yields the distribution function $W_{\nu-i\eta_x} = C
I_{\nu}(\eta_b)$, where $C = (2e I_{i \eta_x}(\eta_b))^{-1}$ is the normalization
constant. Substituting this result into Eq.~(\ref{V}) and using the fact that the
distribution function is real, $W_{-n} = W^*_n$, we obtain, upon integration over $q$,
the following result for the $I$-$V$ characteristics
\begin{equation}
\label{mainresult} \langle V \rangle = I_xR - \frac{k_B T}{e} \frac{\sinh(\pi
\eta_x)}{|I_{i\eta_x}(\eta_b)|^2}.
\end{equation}
Equation (\ref{mainresult}) is the main result of this section. It describes the
complete current-voltage characteristics in the presence of small fluctuations. The
$I$-$V$ characteristics of Eq.~(\ref{mainresult}) are shown in the left panel of
Fig.~\ref{IVcurve} for various values of the parameter $k_BT/eV_b$. For comparison we
plot the current $I$ through the junction as a function of $\langle V \rangle$ in the
right panel of Fig.~\ref{IVcurve}, for the same values of $k_BT/eV_b$. This would
correspond to a measurement in the voltage-biased set-up, see Fig.~\ref{setup}a. We
see that the $I$-$V$ curve corresponding to the limit without fluctuations is
asymptotically reached by the finite temperature $I$-$V$ curves, both for $I_x/I_b \ll
1$ and for $I_x/I_b \gg 1$. Mathematically, this is a direct result of the asymptotics
of the modified Bessel functions~\cite{Gradstein}. Physically, this can be easily
understood from the fact that the point $I_x = I_b$ is unstable, separating a
stationary quasicharge solution from an oscillating one. In the vicinity of $I_b$, any
small current fluctuation will drastically change the nature of the dynamics. Thus the
$I$-$V$ characteristic is strongly affected by fluctuations for $I_x \sim I_b$.

We finally note that the analytical form of the
result~(\ref{mainresult}) resembles the well-known
result~\cite{Likharev86} for the $I$-$V$ characteristic of an
overdamped junction with $R \ll R_Q$ and $E_C \gg E_J$. In fact, these
two cases can be related by duality arguments. Here, we studied the
temperature driven diffusion of quasicharge $q$, governed by
Eq.~(\ref{Fokker}), and found the time-averaged {\em voltage} across
the junction as the average $\langle \sin (\pi q/e) \rangle$ over the
fluctuations of $q$. In an overdamped junction, it is the dual phase
variable $\phi$ that diffuses, governed by an equation similar
to~(\ref{Fokker}) and one is interested in the time averaged {\em
current} $\langle \sin \phi \rangle$ over the fluctuations of
$\phi$. This is why the resulting $I$-$V$ curves can be related to
each other, essentially by exchanging the role of $I$ and $V$.

\section{Discussion}
\label{conclusion} In this paper we have considered the influence of thermal
fluctuations on the current-voltage characteristics of an underdamped Josephson
junction with $R\gg R_Q$ and $E_J \gg E_C$, in the absence of quasiparticles. To
obtain analytical results for the $I$-$V$ characteristics, we have analyzed the
appropriate Langevin equation using a Fokker-Planck approach. The resulting $I$-$V$
characteristic is essentially a Bloch nose~\cite{ALZ85}, smeared by thermal
fluctuations. Our results are valid as long as the inequality~(\ref{Tint}) is
satisfied. Since we restricted our analysis to the lowest energy band we need to
impose in addition $k_B T \ll E_J$ and $I_x \ll I_J$.

The present work is motivated by recent experiments~\cite{Watanabe01}
on a small, voltage biased Josephson junction coupled to a tunable
environment. This environment consisted of a long array with a large
number ($\sim 10^2$) Josephson junctions. Each junction is a small
SQUID-loop, threaded by a magnetic flux $\Phi$. Thus the Josephson
coupling energy $E_J^\mathrm{array}(\Phi)$ of the array is
flux-dependent and the ratio $E_{J}^\mathrm{array}/E_C^\mathrm{array}$
could be changed during the experiment. When $\Phi$ is tuned close to
$\Phi_0/2$, where $\Phi_0$ is the superconducting flux quantum, the
ratio $E_{J}^\mathrm{array}/E_C^\mathrm{array} \ll 1$ and the array is
in the insulating regime.

We ignore quasiparticles: in order for them to reach the small Josephson junction,
they have to tunnel through the entire array; the corresponding probability is
negligibly small. The frequency-dependent impedance $Z(\omega)$ of the array, as seen
by the small junction, is then entirely due to the dynamics of Cooper pairs. In the
stationary limit of interest here, we need to know the zero-frequency component
$Z(\omega=0)$. Calculations by Efetov~\cite{Efetov80} in the limit
$E_{J}^\mathrm{array}/E_C^\mathrm{array} \ll 1$ show that the real part $R$ of
$Z(\omega =0)$ is strictly infinite at zero temperature; at finite $T$, thermal
activation gives rise to a finite, large value of $R$, which depends on the ratio
$E_{J}^\mathrm{array}/E_C^\mathrm{array}$, and thus on $\Phi$. Note that any residual
quasiparticle contribution would be insensitive to the Josephson coupling and hence
independent of $\Phi$. Once $R(\Phi)$ and the parameters of the small junction are
known, our results can be compared in principle with the experimental data
of~\cite{Watanabe01}.\\

\noindent {\bf Acknowledgments}\\

\noindent We thank D.~Averin, K.~Efetov, D. Haviland, E.~Mishchenko, and M. Watanabe
for valuable discussions. One of us (I.~B.) thanks Laboratoire de Physique et
Mod\'elisation des Milieux Condens\'es for hospitality. Our research was sponsored by
Institut Universitaire de France and CNRS-ATIP, Grants DMR-9984002 and BSF-9800338, as
well as by the A.P. Sloan and the Packard Foundations.

\end{document}